\documentclass[preprint,aps,12pt,showpacs,nofootinbib,tightenlines]{revtex4}
\usepackage{amsmath}
\usepackage{amssymb}
\usepackage{epsfig}
\usepackage{graphicx}
\begin{document}
\def\pslash{\rlap{\hspace{0.02cm}/}{p}}
\def\eslash{\rlap{\hspace{0.02cm}/}{e}}
\title{The flavor-changing single-top quark production in the littlest Higgs model with T parity at the LHC}
\author{Xuelei Wang}\email{wangxuelei@sina.com}
\author{Yanju Zhang}
\author{Huiling Jin}
\author{Yanhui Xi}
\affiliation{College of Physics and Information Engineering, Henan
Normal University, Xinxiang, Henan, 453007, P.R. China}
\date{\today}
\begin{abstract}

The littlest Higgs model with discrete symmetry named
"T-parity"(LHT) is an interesting new physics model which does not
suffer strong constraints from electroweak precision data.  One of
the important features of the LHT model is the existence of new
source of FC interactions between the SM fermions and the mirror
fermions. These FC interactions can make significant loop-level
contributions to the couplings $tcV$, and furthermore enhance the
cross sections of the FC single-top quark production processes. In
this paper, we study some FC single-top quark production
processes, $pp\rightarrow t\bar{c}$ and $pp\rightarrow tV$, at the
LHC in the LHT model. We find that the cross sections of these
processes are strongly depended on the mirror quark masses. The
processes $pp\rightarrow t\bar{c}$ and $pp\rightarrow tg$ have
large cross sections with heavy mirror quarks. The observation of
these FC processes at the LHC is certainly the clue of new
physics, and further precise measurements of the cross scetions
can provide useful information about the free parameters in the
LHT model, specially about the mirror quark masses.

\end{abstract}

\pacs{14.65.Ha,12.60.-i, 12.15.Mn,13.85.Lg}

\maketitle


\section{ Introduction}
On the experimental aspect, the forthcoming generation of high
energy colliders, headed by the Large Hadron Collider(LHC) at CERN
depicts an exciting scenario for probing the existence of physics
beyond the Standard Model(SM) of strong and electroweak(EW)
interaction\cite{LHC}. For the probe of new physics at the high
energy colliders like the LHC, there are two ways: one is through
detecting the direct production of new particles and the other is
through unravelling the quantum effects of new physics in some
sensitive and well-measured processes. These two aspects can be
complementary and offer a consistent check for new physics. If the
collider energy is not high enough to produce the heavy new
particles, probing the quantum effects of new particles will be
the only way of peeking at the hints of new physics.

On the other hand, as the heaviest fermion in the SM, the top
quark is speculated to be a sensitive probe of new physics. Due to
the small statistics of the experiments at the Fermilab Tevatron
collider, so far the top quark properties have not been precisely
measured and there remained a plenty of room for new physics
effects in top quark processes. Since the LHC will be a top
factory and allow to scrutinize the top quark nature, unravelling
 new physics effects in various top quark processes will be an
intriguing channel for testing new physics models. Furthermore,
there exists a typical property for the top quark in the SM, i.e.,
its FC interactions are extremely small\cite{FC-SM} due to the
Glashow-Iliopoulos-Maiani(GIM) mechanism. This will make the
observation of any FC top quark process a smoking gun for new
physics. Therefore, the combination of the top quark and FC
processes will be an interesting research field for LHC
experiments.

On the theoretical aspect, the SM is in excellent agreement with
the results
 of particle physics experiments, in particular with the EW
 precision measurements, thus suggesting that the SM cutoff scale
 is at least as large as 10 TeV. Having such a relatively high cutoff,
 however, the SM requires an unsatisfactory fine-tuning to yield a
 correct($\approx 10^2$ GeV) scale for the squared Higgs mass, whose
 corrections are quadratic and therefore highly sensitive to the cutoff.
 This little hierarchy problem has been one of the main
 motivations to elaborate new physics. Recently, an alternative known as
the little Higgs mechanism\cite{little Higgs}, has been proposed.
Such mechanism that makes the Higgs "little" in the current
reincarnation of the PGB idea is collective symmetry breaking.
Collective symmetry breaking protects the Higgs by several
symmetries under each of which the Higgs is an exact Goldstone.
Only if the symmetries are broken collectively, i.e. by more than
one coupling in the theory, can the Higgs pick up a contribution
to its mass and hence all one-loop quadratic divergences to the
Higgs mass are avoided. The most compact implementation of the
little Higgs mechanism is known as the littlest Higgs (LH)
model\cite{LH}. In this model, the SM is enlarged to incorporate
an approximate $SU(5)$ global symmetry. This symmetry is broken
down to $SO(5)$ spontaneously, though the mechanism of this
breaking is left unspecified. The Higgs is an approximate
Goldstone boson of this breaking. In this model there are new
vector bosons, a heavy top quark and a triplet of heavy scalars in
addition to the SM particles. These new particles can make
significant tree-level contributions to the experimental
observables. So the original LH model suffers strong constraints
from electroweak precision data\cite{constraints}. The most
serious constraints result from the tree-level corrections to
precision electroweak observables due to the exchanges of the
additional heavy gauge bosons, as well as from the small but
non-vanishing vacuum expectation value(VEV) of the additional
weak-triplet scalar field.  To solve this problem, a $Z_2$
discrete symmetry named "T-parity" is introduced\cite{LHT}. The
littlest Higgs model with T parity(LHT), requires the introduction
of "mirror fermions" for each SM fermion doublet. The mirror
fermions are odd under T-parity and can be given large masses and
the SM fields are T-even. T parity explicitly forbids any
tree-level contribution from the heavy gauge bosons to the
observables involving only standard model particles as external
states. It also forbids the interactions that induce the triplet
VEV. As a result, in the LHT model, the corrections to the
precision electroweak observables are generated at loop-level.
This implies that the constraints are generically weaker than in
the tree-level case, and fine tuning can be avoided\cite{scale}.

  As we know, there exist
new sources of FC top quark interactions in some new physics
models, such as the Topcolor-assisted Technicolor(TC2) Model and
the Minimal Supersymmetric Standard Model(MSSM). Many studies have
been performed and shown that the existence of FC top quark
interactions in various new physics models can significantly
enhance the branching ratios of the rare top quark
decays\cite{tcv-MSSM,tcv-TC2,tcv-LHT} and the cross sections of
the top-charm production at hadron
colliders\cite{tc-MSSM-hadron,tc-TC2-hadron} and linear
colliders\cite{tc-MSSM-LC, tc-2HDM-LC,tc-TC2-LC,tc-effective-LC}.
Such FC interactions can also significantly influence other FC
processes involving top quark\cite{et-LHT,other FC processes}. Due
to the fact that different models predict different orders of
enhancement, the measurement of these FC top quark processes at
the LHC will provide a unique way to distinguish these models. In
the LHT model, one of the important ingredients of the mirror
sector is the existence of CKM-like unitary mixing matrices. These
mirror mixing matrices parameterize the FC interactions between
the SM fermions and the mirror fermions. Such new FC interactions
also have a very different pattern from ones present in the SM and
can have significant contributions to some FC processes. The
impact of the FC interactions in the LHT on the $K,B,D$ systems
are studied in Refs.\cite{D-LHT1,D-LHT2,D-LHT3}. The FC couplings
between the SM fermions and the mirror fermions can also make the
loop-level contributions to the $tcV(V=\gamma, Z,g)$ couplings.
Such contributions can significantly enhance the branching ratios
of the rare top quark decays $t\rightarrow cV$\cite{tcv-LHT} and
the production rate of the process $eq\rightarrow
et$\cite{et-LHT}. The FC couplings $tcV$ can also make
contributions to the FC top-charm quark production. We have
systematically studied the top-charm quark production at the
International Linear Collider(ILC) and found that these processes
can open an ideal window to probe the LHT model\cite{tc-LHT-LC}.
In this paper, we study the top-charm production at the LHC in the
framework of the LHT model. On the other hand, the single-top
quark can also be produced associated with a neutral gauge boson
via the FC couplings $tcV$, these processes are also studied in
this paper.

\indent  This paper is organized as follows. In Sec.II, we briefly
review the LHT model. Sec.III presents the detailed calculation of
the cross sections for the FC single-top quark production
processes at the LHC. The numerical results are shown in Sec.IV.
We present conclusions and summaries in the last section.

\section{A brief review of the LHT model}

The LH model embeds the electroweak sector of the SM in an
$SU(5)/SO(5)$ non-linear sigma model. It begins with a global
$SU(5)$ symmetry with a locally gauged sub-group $[SU(2)\times
U(1)]^2$. The $SU(5)$ symmetry is spontaneously broken down to
$SO(5)$ via a VEV of order $f$. At the same time, the
$[SU(2)\times U(1)]^2$ gauge symmetry is broken to its diagonal
subgroup $SU(2)_L\times U(1)_Y$ which is identified as the SM
electroweak gauge group. From the $SU(5)/SO(5)$ breaking, there
arise 14 Nambu-Goldstone bosons which are described by the matrix
$\Pi$, given explicitly by
\begin {equation}
\Pi=
\begin{pmatrix}
-\frac{\omega^0}{2}-\frac{\eta}{\sqrt{20}}&-\frac{\omega^+}{\sqrt{2}}
&-i\frac{\pi^+}{\sqrt{2}}&-i\phi^{++}&-i\frac{\phi^+}{\sqrt{2}}\\
-\frac{\omega^-}{\sqrt{2}}&\frac{\omega^0}{2}-\frac{\eta}{\sqrt{20}}
&\frac{v+h+i\pi^0}{2}&-i\frac{\phi^+}{\sqrt{2}}&\frac{-i\phi^0+\phi^P}{\sqrt{2}}\\
i\frac{\pi^-}{\sqrt{2}}&\frac{v+h-i\pi^0}{2}&\sqrt{4/5}\eta&-i\frac{\pi^+}{\sqrt{2}}&
\frac{v+h+i\pi^0}{2}\\
i\phi^{--}&i\frac{\phi^-}{\sqrt{2}}&i\frac{\pi^-}{\sqrt{2}}&
-\frac{\omega^0}{2}-\frac{\eta}{\sqrt{20}}&-\frac{\omega^-}{\sqrt{2}}\\
i\frac{\phi^-}{\sqrt{2}}&\frac{i\phi^0+\phi^P}{\sqrt{2}}&\frac{v+h-i\pi^0}{2}&-\frac{\omega^+}{\sqrt{2}}&
\frac{\omega^0}{2}-\frac{\eta}{\sqrt{20}}
\end{pmatrix}
\end{equation}
Here, $H=(-i\pi^+\sqrt{2},(v+h+i\pi^0)/2)^T$ plays the role of the
SM Higgs doublet, i.e. $h$ is the usual Higgs field, $v=246$ GeV
is the Higgs VEV, and $\pi^{\pm},\pi^0$ are the Goldstone bosons
associated with the spontaneous symmetry breaking $SU(2)_L\times
U(1)_Y\rightarrow U(1)_{em}$. The fields $\eta$ and $\omega$ are
additional Goldstone bosons eaten by heavy gauge bosons when the
$[SU(2)\times U(1)]^2$¡¡gauge group is broken down to
$SU(2)_L\times U(1)_Y$. The field $\Phi$ is a physical scalar
triplet with
\begin {equation}
\Phi=
\begin{pmatrix}
-i\phi^{++}&-i\frac{\phi^+}{\sqrt{2}}\\
-i\frac{\phi^+}{\sqrt{2}}&\frac{-i\phi^0+\phi^P}{\sqrt{2}}
\end{pmatrix}
\end{equation}
Its mass is given by
\begin{eqnarray}
m_{\Phi}=\sqrt{2}m_H\frac{f}{v},
\end{eqnarray}
with $m_H$ being the mass of the SM Higgs scalar.

 In the LHT model, a T-parity discrete symmetry is introduced to make the model consistent
 with the
electroweak precision data. Under the T-parity, the fields
$\Phi,\omega,$ and $\eta$ are odd, and the SM Higgs doublet $H$ is
even.

For the gauge group $[SU(2)\times U(1)]^2$, there are eight gauge
bosons, $W^{a\mu}_1, B^{\mu}_1,W^{a\mu}_2, B^{\mu}_2$(a=1,2,3). A
natural way to define the action of T-parity on the gauge fields
is
\begin{eqnarray}
W^a_1\Leftrightarrow W^a_2,~~~~~~B_1\Leftrightarrow B_2.
\end{eqnarray}
An immediate consequence of this definition is that the gauge
couplings of the two $SU(2)\times U(1)$ factors have to be equal.

The gauge boson T-parity eigenstates are given by
\begin{eqnarray}
W^a_L=\frac{W^a_1+W^a_2}{\sqrt{2}},~~~~~~B_L=\frac{B_1+B_2}{\sqrt{2}}~~~~(T-even)\\
W^a_H=\frac{W^a_1-W^a_2}{\sqrt{2}},~~~~~~B_L=\frac{B_1-B_2}{\sqrt{2}}~~~~(T-odd)
\end{eqnarray}

From the first step of symmetry breaking $[SU(2)\times
U(1)]^2\rightarrow SU(2)_L\times U(1)_Y$, the T-odd heavy gauge
bosons acquire masses. The masses of the T-even gauge bosons are
generated only through the second step of symmetry breaking
$SU(2)_L\times U(1)_Y\rightarrow U(1)_{em}$. Finally, the mass
eigenstates are given at order $O(v^2/f^2)$ by

\begin{eqnarray}
W^{\pm}_L=\frac{W^1_L\mp
iW^2_L}{\sqrt{2}},~~~~~~W^{\pm}_H=\frac{W^1_H\mp
iW^2_H}{\sqrt{2}}\\ \nonumber Z_L=cos\theta_WW^3_L-sin\theta_WB_L,
~~~~~~Z_H=W^3_H+x_H\frac{v^2}{f^2}B_H,\\
\nonumber
A_L=sin\theta_WW^3_L+cos\theta_WB_L,
~~~~~~A_H=-x_H\frac{v^2}{f^2}W^3_H+B_H,
\end{eqnarray}
where $\theta_W$ is the usual weak mixing angle and
\begin{eqnarray}
x_H=\frac{5gg'}{4(5g^2-g'^2)},
\end{eqnarray}
with $g,g'$ being the corresponding coupling constants of
$SU(2)_L$ and $U(1)_Y$. The masses of the T-odd gauge bosons are
given by
\begin{eqnarray}
M_{Z_H}\equiv
M_{W_H}=fg(1-\frac{v^2}{8f^2}),~~~~M_{A_H}=\frac{fg'}{\sqrt{5}}(1-\frac{5v^2}{8f^2}).
\end{eqnarray}
The masses of the T-even gauge bosons are given by
\begin{eqnarray}
M_{W_L}=\frac{gv}{2}(1-\frac{v^2}{12f^2}),
~~~~M_{Z_L}=\frac{gv}{2cos\theta_W}(1-\frac{v^2}{12f^2}),
M_{A_L}=0.
\end{eqnarray}

A consistent and phenomenologically viable implementation of
T-parity in the fermion sector requires the introduction of mirror
fermions. The T-even fermion section consists of the SM quarks,
leptons and an additional heavy quark $T_+$. The T-odd fermion
sector consists of three generations of mirror quarks and leptons
and an additional heavy quark $T_-$. Only the mirror quarks
$(u^i_H,d^i_H)$ are involved in this paper. The mirror quarks get
masses
\begin{eqnarray}
m^u_{H_i}=\sqrt{2}\kappa_if(1-\frac{v^2}{8f^2})\equiv
m_{H_i}(1-\frac{v^2}{8f^2}), \\
\nonumber
m^d_{H_i}=\sqrt{2}\kappa_if\equiv m_{H_i},
\end{eqnarray}
where the Yukawa couplings $\kappa_i$ can in general depend on the
fermion species $i$.

The mirror fermions induce a new flavor structure and there are
four CKM-like unitary mixing matrices in the mirror fermion
sector:
\begin{eqnarray}
V_{H_u},~~V_{H_d},~~V_{H_l},~~V_{H_{\nu}}.
\end{eqnarray}
These mirror mixing matrices are involved in the FC interactions
between the SM fermions and the T-odd mirror fermions which are
mediated by the T-odd heavy gauge bosons or the Goldstone bosons.
$V_{H_u}$ and $V_{H_d}$ satisfy the relation
\begin{eqnarray}
V^{\dag}_{H_u}V_{H_d}=V_{CKM}.
\end{eqnarray}
We parameterize the $V_{H_d}$ with three angles
$\theta^d_{12},\theta^d_{23},\theta^d_{13}$ and three phases
$\delta^d_{12},\delta^d_{23},\delta^d_{13}$

\begin {eqnarray}
V_{H_d}=
\begin{pmatrix}
c^d_{12}c^d_{13}&s^d_{12}s^d_{13}e^{-i\delta^d_{12}}&s^d_{13}e^{-i\delta^d_{13}}\\
-s^d_{12}c^d_{23}e^{i\delta^d_{12}}-c^d_{12}s^d_{23}s^d_{13}e^{i(\delta^d_{13}-\delta^d_{23})}&
c^d_{12}c^d_{23}-s^d_{12}s^d_{23}s^d_{13}e^{i(\delta^d_{13}-\delta^d_{12}-\delta^d_{23})}&
s^d_{23}c^d_{13}e^{-i\delta^d_{23}}\\
s^d_{12}s^d_{23}e^{i(\delta^d_{12}+\delta^d_{23})}-c^d_{12}c^d_{23}s^d_{13}e^{i\delta^d_{13}}&
-c^d_{12}s^d_{23}e^{i\delta^d_{23}}-s^d_{12}c^d_{23}s^d_{13}e^{i(\delta^d_{13}-\delta^d_{12})}&
c^d_{23}c^d_{13}
\end{pmatrix}
\end{eqnarray}
The matrix $V_{H_u}$ is then determined through
$V_{H_u}=V_{H_d}V^{\dag}_{CKM}$. As in the case of the CKM matrix
the angles $\theta^d_{ij}$ can all be made to lie in the first
quadrant with $0\leq
\delta^d_{12},\delta^d_{23},\delta^d_{13}<2\pi$.

\section{The FC single-top quark production processes in the LHT model at the LHC}
\subsection{The loop-level FC couplings $tcV$ in the LHT model}

As we have mentioned above, there are FC interactions between the
SM fermions and the T-odd mirror fermions which are mediated by
the T-odd heavy gauge bosons($A_H,Z_H,W^{\pm}_H$) or Goldstone
bosons($\eta,\omega^0,\omega^{\pm}$). The relevant Feynman rules
can be found in Ref.\cite{D-LHT1}. With these FC couplings, the
loop-level FC couplings $tcV$  can be induced and the relevant
Feynman diagrams are shown in Fig.1.

As we know, each diagram in Fig.1 actually contains ultraviolet
divergence. Because there is no corresponding tree-level $tcV$
coupling to absorb these divergences, the divergences just cancel
each other and the total effective $tcV$ couplings are finite as
they should be. The effective one loop-level couplings $tcV$
 can be directly calculated based on
Fig.1. Their explicit forms, $\Gamma^{\mu}_{tc\gamma}(p_t,p_c)$,
$\Gamma^{\mu}_{tcZ}(p_t,p_c)$ and $\Gamma^{\mu}_{tcg}(p_t,p_c)$,
are given in Appendix.

With the FC couplings $tcV$, the top-charm quarks can be produced
via gluon-gluon collision or $q\bar{q}$ collision. On the other
hand, single-top quark can also be produced associated with a SM
gauge boson via charm-gluon collision. We will study these
processes in the following.

\subsection{The top-charm quark production in the LHT model at the LHC}

 \hspace{1mm}
In the LHT model, the existence of the FC couplings $tcV$ can
induce the subprocesses $gg\rightarrow t\bar{c}$ and
$q\bar{q}\rightarrow t\bar{c}$ at loop-level. The corresponding
Feynman diagrams are shown in Fig.2, and the production amplitudes
are

\begin{eqnarray}
M_{A}&=&ig_{s}f^{abc}G(p_1+p_2)\bar{u}^i_{t}(p_{3})\Gamma^{\mu
aij}_{tcg}(p_{3},-p_{4})[(p_{1}-p_{2})_{\mu}
\epsilon^c(p_1)\cdot\epsilon^b(p_2)\\ \nonumber &&
+2p_{2}\cdot\epsilon^c(p_1)\epsilon_{\mu}^b(p_2)
-2p_{1}\cdot\epsilon^b(p_2)\epsilon_{\mu}^c(p_1)]v^j_{\bar{c}}(p_{4}),
\end{eqnarray}
\begin{eqnarray}
 M_{B}&=&-g_{s}T^{bjk}G(p_3-p_1,m_c)\bar{u}^i_{t}(p_{3})\Gamma^{\mu aij}_{tcg}(p_{3},p_{3}-p_{1})\epsilon^a_{\mu}(p_{1})
 \\ \nonumber
 &&(\pslash_3-\pslash_1+m_c)\rlap/\epsilon^b(p_2)v^k_{\bar{c}}(p_{4}),
 \end{eqnarray}
\begin{eqnarray}
 M_{C}&=&-g_{s}T^{aij}G(p_3-p_1,m_t)\bar{u}^i_{t}(p_{3})\rlap/\epsilon^a(p_1)(\pslash_3-\pslash_1+m_{t})
 \\ \nonumber
 &&\Gamma^{\mu bjk}_{tcg}(p_3-p_1,-p_{4})\epsilon^b_{\mu}(p_2)
 v^k_{\bar{c}}(p_{4}),
 \end{eqnarray}
\begin{eqnarray}
M_{D}&=&g_{s}T^{alj}G(p_1+p_2,0)\bar{u}^i_{t}(p_{3})\Gamma^{\mu
  aik}_{tcg}(p_{3},-p_{4})v^k_{\bar{c}}(p_{4})\bar{v}^l_{\bar{q}}(p_{2})
\gamma_{\mu}u^j_{q}(p_{1}),
\end{eqnarray}
\begin{eqnarray}
M_{E}&=&g_{s}T^{alk}G(p_3-p_1,0)\bar{u}^i_{t}(p_{3})\Gamma^{\mu
aij}_{tcg}(p_{3},p_{1})u^j_{c}(p_{1})\bar{v}^l_{\bar{c}}(p_{2})
\gamma_{\mu}v^k_{\bar{c}}(p_{4}).
\end{eqnarray}

Here $p_{1},p_{2}$ are the momenta of the incoming states, and
$p_{3},p_{4}$ are the momenta of the outgoing final states top
quark and anti-charm quark, respectively. We also define $G(p, m)$
as $\frac{1}{p^2-m^2}$.

\subsection{The $tV$ production in the LHT model at the LHC}

The FC couplings $tcV$ can also induce the FC single-top quark
production $cg\rightarrow tV$ at hadron colliders. The
corresponding Feynman diagrams are shown in Fig.3, and the
production amplitudes can be written as

\begin{eqnarray}
M_{F}^{\gamma}&=&-\frac{2e}{3}G(p_1+p_2,m_t)\bar{u}^i_{t}(p_{3})\rlap/\epsilon(p_4)
(\pslash_1+\pslash_2+m_{t}) \Gamma^{\mu
aij}_{tcg}(p_{1}+p_{2},p_{1})\epsilon_{\mu}(p_{2})u^j_{c}(p_{1}),
\end{eqnarray}
\begin{eqnarray}
M_{F}^{Z}&=&-\frac{g}{\cos\theta_{W}}G(p_1+p_2,m_t)\bar{u}^i_{t}(p_{3})\rlap/\epsilon(p_4)[(\frac{1}{2}-\frac{2}{3}\sin^{2}\theta_{W})P_{L}
-\frac{2}{3}\sin^{2}\theta_{W}P_{R}]
\\ \nonumber
&&(\pslash_1+\pslash_2+m_{t})\times\Gamma^{\mu
aij}_{tcg}(p_{1}+p_{2},p_{1})\epsilon_{\mu}(p_{2})u^j_{c}(p_{1}),
\end{eqnarray}
\begin{eqnarray}
M_{F}^{g}&=&-g_{s}T^{bil}G(p_1+p_2,m_t)\bar{u}^i_{t}(p_{3})\rlap/\epsilon^b(p_4)(\pslash_1+\pslash_2+m_{t})\\
 \nonumber
&&\Gamma^{\mu
alj}_{tcg}(p_{1}+p_{2},p_{1})\epsilon^a_{\mu}(p_{2})u^j_{c}(p_{1}),
\end{eqnarray}
\begin{eqnarray}
M_{G}^{\gamma}&=&-g_{s}T^{aij}G(p_1+p_2,m_{c})\bar{u}^i_{t}(p_{3})\Gamma^{\mu}_{tc\gamma}(p_{3},p_{3}+p_{4})\\
\nonumber
&&\epsilon_{\mu}(p_{4})(\pslash_1+\pslash_2+m_{c})\rlap/\epsilon^a(p_2)u^j_{c}(p_{1}),
\end{eqnarray}
\begin{eqnarray}
M_{G}^{Z}&=&-g_{s}T^{aij}G(p_1+p_2,m_{c})\bar{u}^i_{t}(p_{3})\Gamma^{\mu}_{tcZ}(p_{3},p_{3}+p_{4})
\\
\nonumber
&&\epsilon_{\mu}(p_{4})(\pslash_1+\pslash_2+m_{c})\rlap/\epsilon^{a}(p_2)u^{j}_{c}(p_{1}),
\end{eqnarray}
\begin{eqnarray}
M_{G}^{g}&=&-g_{s}T^{alj}G(p_1+p_2,m_{c})\bar{u}^i_{t}(p_{3})\Gamma^{\mu
bil}_{tcg}(p_{3},p_{3}+p_{4})
\\
\nonumber
&&\epsilon^b_{\mu}(p_{4})(\pslash_1+\pslash_2+m_{c})\rlap/\epsilon^a(p_2)u^j_{c}(p_{1}),
\end{eqnarray}
\begin{eqnarray}
M_{H}^{\gamma}&=&-g_{s}T^{aij}G(p_3-p_2,m_t)\bar{u}^i_{t}(p_{3})\rlap/\epsilon^a(p_2)(\pslash_3-\pslash_2+m_{t})
\\
\nonumber
&&\Gamma^{\mu}_{tc\gamma}(p_{1}-p_4,p_{1})\epsilon_{\mu}(p_{4})u^j_{c}(p_{1}),
\end{eqnarray}
\begin{eqnarray}
M_{H}^{Z}&=&-g_{s}T^{aij}G(p_3-p_2,m_t)\bar{u}^i_{t}(p_{3})\rlap/\epsilon^a(p_2)(\pslash_3-\pslash_2+m_{t})
\\
\nonumber
&&\Gamma^{\mu}_{tcZ}(p_1-p_4,p_{1})\epsilon_{\mu}(p_{4})u^j_{c}(p_{1}),
\end{eqnarray}
\begin{eqnarray}
M_{H}^{g}&=&-g_{s}T^{ail}G(p_3-p_2,m_t)\bar{u}^i_{t}(p_{3})\rlap/\epsilon^a(p_2)(\pslash_3-\pslash_2+m_{t})
\\
\nonumber &&\Gamma^{\mu
blj}_{tcg}(p_1-p_4,p_{1})\epsilon^b_{\mu}(p_{4})u^j_{c}(p_{1}),
\end{eqnarray}
\begin{eqnarray}
M_{I}^{\gamma}&=&-\frac{2e}{3}G(p_3-p_2,m_c)\bar{u}^i_{t}(p_{3})\Gamma^{\mu
aij}_{tcg}(p_{3},p_3-p_2)
\epsilon^a_{\mu}(p_{2})(\pslash_3-\pslash_2+m_c)\rlap/\epsilon(p_4)u^j_{c}(p_{1}),
\end{eqnarray}
\begin{eqnarray}
M_{I}^{Z}&=&-\frac{g}{\cos\theta_{W}}G(p_3-p_2,m_c)\bar{u}^i_{t}(p_{3})\Gamma^{\mu
aij}_{tcg}(p_{3},p_3-p_2)
\epsilon^a_{\mu}(p_{2})(\pslash_3-\pslash_2+m_c)\rlap/\epsilon(p_4)\\
\nonumber
&&\times[(\frac{1}{2}-\frac{2}{3}\sin^{2}\theta_{W})P_{L}-\frac{2}{3}\sin^{2}\theta_{W}P_{R}]u^j_{c}(p_{1}),
\end{eqnarray}
\begin{eqnarray}
M_{I}^{g}&=&-g_{s}T^{blj}G(p_3-p_2,m_c)\bar{u}^i_{t}(p_{3})\Gamma^{\mu
ail}_{tcg}(p_3,p_3-p_2)
\\
\nonumber
&&\epsilon^a_{\mu}(p_{2})(\pslash_3-\pslash_2+m_c)\rlap/\epsilon^b(p_4)u^j_{c}(p_{1}).
\end{eqnarray}

With the above production amplitudes, we can directly obtain the
cross sections $\hat{\sigma}_{ij}(\hat{s})$ of the subprocesses
$gg\rightarrow t\bar{c}$, $q\bar{q}\rightarrow t\bar{c}$ and
$cg\rightarrow tV$, where $\hat{s}=(p_{1}+p_{2})^{2}$. The
hadronic cross sections at the hadron colliders can be obtained by
folding the cross sections of the subprocesses with the parton
distribution functions:$f_i^A(x_1,Q)$and $f_j^B(x_2,Q)$, which is
given by
\begin{eqnarray}
\sigma(s)&=&\sum_{ij}\int
dx_{1}dx_{2}[f_i^A(x_1,Q)f_j^B(x_2,Q)+f_i^B(x_1,Q)f_j^A(x_2,Q)]\hat{\sigma}^{ij}(\hat{s},\alpha_{s}(\mu)).
\end{eqnarray}
Thereinto, $Q$ is the factorization scale, $\mu$ is the
renormalization scale, $\sqrt{s}$ is the center-of-mass(c.m.)
energy of the hadron colliders. Here we used the parton
distribution functions that were given by CTEQ6L.

To obtain numerical results of the cross sections, we calculate
the amplitudes numerically by using the method of
reference\cite{HZ}, instead of calculating the square of the
production amplitudes analytically. This greatly simplifies our
calculations.

\section{The numerical results and discussions}

There are several free parameters in the LHT model which are
involved in the production amplitudes. They are the breaking scale
$f$, the mirror quark masses $m_{H_i}(i=1,2,3)$(Here we have
ignored the mass difference between the up-type mirror quarks and
the down-type mirror quarks), and 6
parameters($\theta^d_{12},~\theta^d_{13},~\theta^d_{23},~\delta^d_{12},~\delta^d_{13},~\delta^d_{23}$)
which are related to the mixing matrix $V_{H_d}$. In
Refs.\cite{D-LHT1,D-LHT2,D-LHT3}, the constraints on the mass
spectrum of the mirror fermions have been investigated from the
analysis of neutral meson mixing in the $K,~B$ and $D$ systems.
They found that a TeV scale GIM suppression is necessary for a
generic choice of $V_{H_d}$. However, there are regions of
parameter space where are only very loose constraints on the mass
spectrum of the mirror fermions. Here we calculate the cross
sections based on the two scenarios for the structure of the
matrix $V_{H_d}$, as in Ref.\cite{tcv-LHT}. i.e.,

\hspace{1cm} Case I: $V_{H_d}=1,~~~$$V_{H_u}=V^{\dag}_{CKM}$,

\hspace{1cm} Case II:
$s^d_{23}=1/\sqrt{2},~~s^d_{12}=s^d_{13}=0,~~\delta^d_{12}=\delta^d_{23}=\delta^d_{13}=0.$

In both cases, the constraints on the mass spectrum of the mirror
fermions are very relaxed. For the breaking scale $f$, we take two
typical values: 500 GeV and 1000 GeV.

To get the numerical results of the cross sections, we should also
fix some parameters in the SM as $m_{t}=$174.2 GeV, $m_{c}=$1.25
GeV, $s^{2}_{W}=$0.23, $M_{Z}=$91.87 GeV, $\alpha_e=1/128$,
$\alpha_s=0.1$, and $v=246$ GeV\cite{parameters}. On the other
hand, taking account of the detector acceptance, we have taken the
basic cuts on the transverse momenta($p_{T}$) and the
pseudo-rapidities($\eta$) for the final state particles
\begin{eqnarray*}
p_{T}\geq15 GeV, \hspace{1cm}|\eta|\leq 2.5.
\end{eqnarray*}

The numerical results of the cross sections for the FC single-top
quark production processes at the LHC are summarized in Figs.4-5,
and here the anti-top quark($\bar{t}$) production is also included
in our calculation. The numerical results for Case I are shown in
Fig.4. In case I, the mixing in the down type gauge and Goldstone
boson interactions is absent. In this case, there are no
constraints on the mirror quark masses at one loop-level from the
$K$ and $B$ systems and the constraints come only from the $D$
system. The constraints on the mass of the third generation mirror
quark are very weak. Here, we take $M_{H_3}$ to vary in the range
from 500 GeV to 5000 GeV, and fix $m_{H1}=m_{H2}=300$ GeV. To see
the influence of the scale $f$ on the cross sections, we also take
$f=500,~1000$ GeV, respectively. We can see from Fig.4 that all
the cross sections of the FC single-top quark production processes
rise very fast with the $m_{H_3}$ increasing. This is because the
couplings between the mirror quarks and the SM quarks are
proportion to the mirror quark masses. Among all the single-top
quark production processes,  the process $pp\rightarrow t\bar{c}$
possesses the largest cross section. For the heavy mirror quarks,
the cross section of $pp\rightarrow t\bar{c}$ can reach the level
of a few pb. The cross sections of $pp\rightarrow t\gamma(Z)$ are
much smaller than that of $pp\rightarrow t\bar{c}$ and their cross
sections can only reach a few fb with relative large value of
mirror quark masses. On the other hand, we find that the process
$pp\rightarrow tg$ can also have a sizeable cross section and a
large number of $tg$ events can be produced at the LHC. The scale
$f$ is insensitive to the cross sections. The reason is that the
masses of the heavy gauge bosons and the mirror quarks,
$M_{V_{H}}$ and $m_{H_i}$, are proportion to $f$ but the
production amplitudes are represented in the form of
$m_{H_i}/M_{V_{H}}$ which cancels the effect of $f$. For Case II,
the dependence of the cross sections on $m_{H_3}$ is presented in
Fig.5. In this case, the constraints from the $K$ and B systems
are also very weak. Compared to Case I, the mixing between the
second and third generations is enhanced with the choice of a
bigger mixing angle $s^d_{23}$. The dependence of the free
parameters on the cross sections is similar to that in Case I.
Even with stricter constraints on the mirror quark masses, the
cross sections of the FC single-top quark production processes can
also reach a sizeable level. Specially, the processes
$pp\rightarrow t\bar{c}$ and $pp\rightarrow tg$ benefit from their
large cross sections.

\section{Conclusions and summaries}
\hspace{1mm} In this paper, we study some interesting FC
single-top quark production processes, $pp\rightarrow t\bar{c}$
and $pp\rightarrow tV$, at the LHC in the framework of the LHT
model. We can conclude that (1)All the cross sections of the FC
single-top quark processes are strongly depended on the mirror
quark masses and the cross sections increase sharply with the
mirror quark masses increasing. (2)The cross sections are
insensitive to the scale $f$. (3)The cross section of the process
$pp\rightarrow t\bar{c}$ is the largest one and its cross section
can reach a few pb. The process $pp\rightarrow tg$ also has a
sizeable cross section but the cross sections of $pp\rightarrow
t\gamma (Z)$ are much smaller. With the running of the LHC, it
should have a powerful ability to probe the LHT model via the FC
single-top quark production.

\section{Acknowledgments}
\hspace{1mm}

We would thank Junjie Cao for useful discussions and providing the
calculation programs. This work is supported by the National
Natural Science Foundation of China under Grant No.10775039,
10575029 and 10505007.

 \newpage
 \begin{center}
   \Large{Appendix: The explicit expressions of the effective $tcV$ couplings}
\end{center}

The effective $tcV$ couplings
$\Gamma^{\mu}_{tc\gamma},~\Gamma^{\mu}_{tcZ},~\Gamma^{\mu}_{tcg}$
can be directly calculated based on Fig.1, and they can be
represented  in form of 2-point and 3-point standard functions
$B_0,B_1,C_{ij}$. Due to $m_t>>m_c$, we have safely ignored the
terms $m_c/m_t$ in the calculation. On the other hand, the high
order $1/f^2$ terms in the masses of new gauge bosons and in the
Feynman rules are also ignored.
$\Gamma^{\mu}_{tc\gamma},~\Gamma^{\mu}_{tcZ},~\Gamma^{\mu}_{tcg}$
are depended on the momenta of top quark and charm
quark($p_t,p_c$). Here $p_t$ is outgoing and $p_c$ is incoming.
The explicit expressions of them are

\begin{eqnarray*}
\Gamma^{\mu aij}_{tcg}(p_t,p_c)&=&\Gamma^{\mu
aij}_{tcg}(\eta^{0})+\Gamma^{\mu aij}_{tcg}(\omega^{0})
+\Gamma^{\mu aij}_{tcg}(\omega^{\pm})+\Gamma^{\mu
aij}_{tcg}(A_{H})+\Gamma^{\mu aij}_{tcg}(Z_{H}) +\Gamma^{\mu
aij}_{tcg}(W_{H}^{\pm}).
\end{eqnarray*}
\begin{eqnarray*}
\Gamma^{\mu
aij}_{tcg}(\eta^{0})&=&\frac{i}{16\pi^{2}}\frac{g^{\prime2}}{100M_{A_{H}}^{2}}
(V_{Hu})_{3i}(V_{Hu})_{i2}{m_{Hi}^{2}}g_{s}T^{aij}\\&&
\{[B_{0}(-p_{t},m_{Hi},0)-B_{0}(-p_{c},m_{Hi},0)
+B_{1}(-p_{t},m_{Hi},0)\\&&+2C_{24}^{a} -2p_{t}\cdot
p_{c}(C_{12}^{a}+C_{23}^{a})+m_{t}^{2}(C_{21}^{a}+C_{11}^{a}+C_{0}^{a})-m_{Hi}^{2}C_{0}^{a}]
\gamma^{\mu}P_{L}\\&&
+[-2m_{t}(C_{21}^{a}+2C_{11}^{a}+C_{0}^{a})]p_{t}^{\mu}P_{L}+[2m_{t}(C_{23}^{a}+2C_{12}^{a})]p_{c}^{\mu}P_{L}\},
\end{eqnarray*}
\begin{eqnarray*}
\Gamma^{\mu
aij}_{tcg}(\omega^{0})&=&\frac{i}{16\pi^{2}}\frac{g^{2}}{4M_{Z_{H}}^{2}}
(V_{Hu})_{3i}(V_{Hu})_{i2}{m_{Hi}^{2}}g_{s}T^{aij}\\&&
\{[B_{0}(-p_{t},m_{Hi},0)-B_{0}(-p_{c},m_{Hi},0)
+B_{1}(-p_{t},m_{Hi},0)\\&&+2C_{24}^{b} -2p_{t}\cdot
p_{c}(C_{12}^{b}+C_{23}^{b})+m_{t}^{2}(C_{21}^{b}+C_{11}^{b}+C_{0}^{b})-m_{Hi}^{2}C_{0}^{b}]
\gamma^{\mu}P_{L}\\&&
+[-2m_{t}(C_{21}^{b}+2C_{11}^{b}+C_{0}^{b})]p_{t}^{\mu}P_{L}+[2m_{t}(C_{23}^{b}+2C_{12}^{b})]p_{c}^{\mu}P_{L}\},
\end{eqnarray*}
\begin{eqnarray*}
\Gamma^{\mu
aij}_{tcg}(\omega^{\pm})&=&\frac{i}{16\pi^{2}}\frac{g^{2}}{2M_{W_{H}}^{2}}
(V_{Hu})_{3i}(V_{Hu})_{i2}{m_{Hi}^{2}}g_{s}T^{aij}\\&&
\{[B_{0}(-p_{t},m_{Hi},0)-B_{0}(-p_{c},m_{Hi},0)
+B_{1}(-p_{t},m_{Hi},0)\\&&+2C_{24}^{c} -2p_{t}\cdot
p_{c}(C_{12}^{c}+C_{23}^{c})+m_{t}^{2}(C_{21}^{c}+C_{11}^{c}+C_{0}^{c})-m_{Hi}^{2}C_{0}^{c}]
\gamma^{\mu}P_{L}\\&&
+[-2m_{t}(C_{21}^{c}+2C_{11}^{c}+C_{0}^{c})]p_{t}^{\mu}P_{L}+[2m_{t}(C_{23}^{c}+2C_{12}^{c})]p_{c}^{\mu}P_{L}\},
\end{eqnarray*}
\begin{eqnarray*}
\Gamma^{\mu
aij}_{tcg}(A_{H})&=&\frac{i}{16\pi^{2}}\frac{g^{\prime2}}{50}
(V_{Hu})_{3i}(V_{Hu})_{i2}g_{s}T^{aij}\\&&
\{[B_{1}(-p_{t},m_{Hi},M_{A_{H}})+2C_{24}^{d} -2p_{t}\cdot
p_{c}(C_{11}^{d}+C_{23}^{d})+m_{t}^{2}(C_{21}^{d}+C_{11}^{d})\\&&-m_{Hi}^{2}C_{0}^{d}]
\gamma^{\mu}P_{L}
+[-2m_{t}(C_{21}^{d}+C_{11}^{d})]p_{t}^{\mu}P_{L}
+[2m_{t}(C_{23}^{d}+C_{11}^{d})]p_{c}^{\mu}P_{L}\},
\end{eqnarray*}
\begin{eqnarray*}
\Gamma^{\mu aij}_{tcg}(Z_{H})&=&\frac{i}{16\pi^{2}}\frac{g^{2}}{2}
(V_{Hu})_{3i}(V_{Hu})_{i2}g_{s}T^{aij}\\&&
\{[B_{1}(-p_{t},m_{Hi},M_{Z_{H}})+2C_{24}^{e} -2p_{t}\cdot
p_{c}(C_{11}^{e}+C_{23}^{e})+m_{t}^{2}(C_{21}^{e}+C_{11}^{e})\\&&-m_{Hi}^{2}C_{0}^{e}]
\gamma^{\mu}P_{L}
+[-2m_{t}(C_{21}^{e}+C_{11}^{e})]p_{t}^{\mu}P_{L}+[2m_{t}(C_{23}^{e}+C_{11}^{e})]p_{c}^{\mu}P_{L}\},
\end{eqnarray*}
\begin{eqnarray*}
\Gamma^{\mu aij}_{tcg}(W_{H}^{\pm})&=&\frac{i}{16\pi^{2}}{g^{2}}
(V_{Hu})_{3i}(V_{Hu})_{i2}g_{s}T^{aij}\\&&
\{[B_{1}(-p_{t},m_{Hi},M_{W_{H}})+2C_{24}^{f} -2p_{t}\cdot
p_{c}(C_{11}^{f}+C_{23}^{f})+m_{t}^{2}(C_{21}^{f}+C_{11}^{f})\\&&-m_{Hi}^{2}C_{0}^{f}]
\gamma^{\mu}P_{L}
+[-2m_{t}(C_{21}^{f}+C_{11}^{f})]p_{t}^{\mu}P_{L}+[2m_{t}(C_{23}^{f}+C_{11}^{f})]p_{c}^{\mu}P_{L}\}.
\end{eqnarray*}

\begin{eqnarray*}
\Gamma^{\mu}_{tc\gamma}(p_t,p_c)&=&\Gamma^{\mu}_{tc\gamma}(\eta^{0})+\Gamma^{\mu}_{tc\gamma}(\omega^{0})
+\Gamma^{\mu}_{tc\gamma}(\omega^{\pm})+\Gamma^{\mu}_{tc\gamma}(A_{H})+\Gamma^{\mu}_{tc\gamma}(Z_{H})
+\Gamma^{\mu}_{tc\gamma}(W_{H}^{\pm})\\&&+\Gamma^{\mu}_{tc\gamma}(W_{H}^{\pm}\omega^{\pm}),
\end{eqnarray*}
\begin{eqnarray*}
\Gamma^{\mu}_{tc\gamma}(\eta^{0})&=&\frac{i}{16\pi^{2}}\frac{eg^{\prime2}}{150M_{A_{H}}^{2}}
(V_{Hu})_{3i}(V_{Hu})_{i2}{m_{Hi}^{2}}\\&&
\{[B_{0}(-p_{t},m_{Hi},0)-B_{0}(-p_{c},m_{Hi},0)
+B_{1}(-p_{t},m_{Hi},0)\\&&+2C_{24}^{a} -2p_{t}\cdot
p_{c}(C_{12}^{a}+C_{23}^{a})+m_{t}^{2}(C_{21}^{a}+C_{11}^{a}+C_{0}^{a})-m_{Hi}^{2}C_{0}^{a}]
\gamma^{\mu}P_{L}\\&&
+[-2m_{t}(C_{21}^{a}+2C_{11}^{a}+C_{0}^{a})]p_{t}^{\mu}P_{L}+[2m_{t}(C_{23}^{a}+2C_{12}^{a})]p_{c}^{\mu}P_{L}\},
\end{eqnarray*}
\begin{eqnarray*}
\Gamma^{\mu}_{tc\gamma}(\omega^{0})&=&\frac{i}{16\pi^{2}}\frac{eg^{2}}{6M_{Z_{H}}^{2}}
(V_{Hu})_{3i}(V_{Hu})_{i2}{m_{Hi}^{2}}\\&&
\{[B_{0}(-p_{t},m_{Hi},0)-B_{0}(-p_{c},m_{Hi},0)
+B_{1}(-p_{t},m_{Hi},0)\\&&+2C_{24}^{b} -2p_{t}\cdot
p_{c}(C_{12}^{b}+C_{23}^{b})+m_{t}^{2}(C_{21}^{b}+C_{11}^{b}+C_{0}^{b})-m_{Hi}^{2}C_{0}^{b}]
\gamma^{\mu}P_{L}\\&&
+[-2m_{t}(C_{21}^{b}+2C_{11}^{b}+C_{0}^{b})]p_{t}^{\mu}P_{L}+[2m_{t}(C_{23}^{b}+2C_{12}^{b})]p_{c}^{\mu}P_{L}\},
\end{eqnarray*}
\begin{eqnarray*}
\Gamma^{\mu}_{tc\gamma}(\omega^{\pm})&=&\frac{i}{16\pi^{2}}\frac{eg^{2}}{6M_{W_{H}}^{2}}
(V_{Hu})_{3i}(V_{Hu})_{i2}{m_{Hi}^{2}}\\&&
\{2[(B_{0}(-p_{t},m_{Hi},0)-B_{0}(-p_{c},m_{Hi},0)
+B_{1}(-p_{t},m_{Hi},0))\\&&-2C_{24}^{c}+6C_{24}^{g}+2p_{t}\cdot
p_{c} (C_{12}^{c}+C_{23}^{c})-m_{t}^{2}(C_{21}^{c}
+C_{11}^{c}+C_{0}^{c})+m_{Hi}^{2}C_{0}^{c}] \gamma^{\mu}P_{L}\\&&
+[2m_{t}(C_{21}^{c}+2C_{11}^{c}+C_{0}^{c})+3m_{t}(2C_{21}^{g}+C_{11}^{g})]p_{t}^{\mu}P_{L}\\&&
+[-2m_{t}(C_{23}^{c}+2C_{12}^{c})-3m_{t}(2C_{23}^{g}+C_{11}^{g})]p_{c}^{\mu}P_{L}\},
\end{eqnarray*}
\begin{eqnarray*}
\Gamma^{\mu}_{tc\gamma}(A_{H})&=&\frac{i}{16\pi^{2}}\frac{eg^{\prime2}}{75}
(V_{Hu})_{3i}(V_{Hu})_{i2}\\&&
\{[B_{1}(-p_{t},m_{Hi},M_{A_{H}})+2C_{24}^{d} -2p_{t}\cdot
p_{c}(C_{11}^{d}+C_{23}^{d})+m_{t}^{2}(C_{21}^{d}+C_{11}^{d})\\&&-m_{Hi}^{2}C_{0}^{d}]
\gamma^{\mu}P_{L}
+[-2m_{t}(C_{21}^{d}+C_{11}^{d})]p_{t}^{\mu}P_{L}
+[2m_{t}(C_{23}^{d}+C_{11}^{d})]p_{c}^{\mu}P_{L}\},
\end{eqnarray*}
\begin{eqnarray*}
\Gamma^{\mu}_{tc\gamma}(Z_{H})&=&\frac{i}{16\pi^{2}}\frac{eg^{2}}{3}
(V_{Hu})_{3i}(V_{Hu})_{i2}\\&&
\{[B_{1}(-p_{t},m_{Hi},M_{Z_{H}})+2C_{24}^{e} -2p_{t}\cdot
p_{c}(C_{11}^{e}+C_{23}^{e})+m_{t}^{2}(C_{21}^{e}+C_{11}^{e})\\&&-m_{Hi}^{2}C_{0}^{e}]
\gamma^{\mu}P_{L}
+[-2m_{t}(C_{21}^{e}+C_{11}^{e})]p_{t}^{\mu}P_{L}
+[2m_{t}(C_{23}^{e}+C_{11}^{e})]p_{c}^{\mu}P_{L}\},
\end{eqnarray*}
\begin{eqnarray*}
\Gamma^{\mu}_{tc\gamma}(W_{H}^{\pm})&=&\frac{i}{16\pi^{2}}\frac{eg^{2}}{6}
(V_{Hu})_{3i}(V_{Hu})_{i2}\\&&
\{[4B_{1}(-p_{t},m_{Hi},M_{W_{H}})+2B_{0}(p_{c},m_{Hi},M_{W_{H}})-4C_{24}^{f}
+4C_{24}^{h}\\&&+4p_{t}\cdot
p_{c}(C_{11}^{f}+C_{23}^{f})-2m_{t}^{2}(C_{21}^{f}+C_{11}^{f})+2m_{Hi}^{2}C_{0}^{f}
+2M_{W_{H}}^{2}C_{0}^{h}\\&&-4p_{t}\cdot
p_{c}(C_{11}^{h}+C_{0}^{h})+m_{t}^{2}(3C_{11}^{h}+C_{0}^{h})]
\gamma^{\mu}P_{L}\\&&
+[4m_{t}(C_{21}^{f}+C_{11}^{f})+2m_{t}(3C_{11}^{h}+2C_{21}^{h}+C_{0}^{h})]p_{t}^{\mu}P_{L}\\&&
+[-4m_{t}(C_{23}^{f}+C_{11}^{f})-2m_{t}(2C_{23}^{h}+3C_{12}^{h}-C_{11}^{h}-C_{0}^{h})]p_{c}^{\mu}P_{L}\},
\end{eqnarray*}
\begin{eqnarray*}
\Gamma^{\mu}_{tc\gamma}(W_{H}^{\pm}\omega^{\pm})&=&\frac{i}{16\pi^{2}}\frac{eg^{2}}0{2}
(V_{Hu})_{3i}(V_{Hu})_{i2}\\&&\{[m_{Hi}^{2}(C_{0}^{i}-C_{0}^{j})+m_{t}^{2}(C_{11}^{j}+C_{0}^{j})]
\gamma^{\mu}P_{L}+[-2m_{t}C_{12}^{j}]p_{c}^{\mu}P_{L}\}.
\end{eqnarray*}
\begin{eqnarray*}
\Gamma^{\mu}_{tcZ}(p_t,p_c)&=&\Gamma^{\mu}_{tcZ}(\eta^{0})+\Gamma^{\mu}_{tcZ}(\omega^{0})
+\Gamma^{\mu}_{tcZ}(\omega^{\pm})+\Gamma^{\mu}_{tcZ}(A_{H})+\Gamma^{\mu}_{tcZ}(Z_{H})
+\Gamma^{\mu}_{tcZ}(W_{H}^{\pm})\\&&+\Gamma^{\mu}_{tcZ}(W_{H}^{\pm}\omega^{\pm}),
\end{eqnarray*}
\begin{eqnarray*}
\Gamma^{\mu}_{tcZ}(\eta^{0})&=&\frac{i}{16\pi^{2}}\frac{g}{\cos\theta_{W}}
(\frac{1}{2}-\frac{2}{3}\sin^{2}\theta_{W})\frac{g^{\prime2}}{100M_{A_{H}}^{2}}
(V_{Hu})_{3i}(V_{Hu})_{i2}{m_{Hi}^{2}}\\&&
\{[B_{0}(-p_{t},m_{Hi},0)-B_{0}(-p_{c},m_{Hi},0)
+B_{1}(-p_{t},m_{Hi},0)\\&&+2C_{24}^{a} -2p_{t}\cdot
p_{c}(C_{12}^{a}+C_{23}^{a})+m_{t}^{2}(C_{21}^{a}+C_{11}^{a}+C_{0}^{a})-m_{Hi}^{2}C_{0}^{a}]
\gamma^{\mu}P_{L}\\&&
+[-2m_{t}(C_{21}^{a}+2C_{11}^{a}+C_{0}^{a})]p_{t}^{\mu}P_{L}+[2m_{t}(C_{23}^{a}+2C_{12}^{a})]p_{c}^{\mu}P_{L}\},
\end{eqnarray*}
\begin{eqnarray*}
\Gamma^{\mu}_{tcZ}(\omega^{0})&=&\frac{i}{16\pi^{2}}\frac{g}{\cos\theta_{W}}
(\frac{1}{2}-\frac{2}{3}\sin^{2}\theta_{W})\frac{g^{2}}{4M_{Z_{H}}^{2}}
(V_{Hu})_{3i}(V_{Hu})_{i2}{m_{Hi}^{2}}\\&&
\{[B_{0}(-p_{t},m_{Hi},0)-B_{0}(-p_{c},m_{Hi},0)
+B_{1}(-p_{t},m_{Hi},0)\\&&+2C_{24}^{b} -2p_{t}\cdot
p_{c}(C_{12}^{b}+C_{23}^{b})+m_{t}^{2}(C_{21}^{b}+C_{11}^{b}+C_{0}^{b})-m_{Hi}^{2}C_{0}^{b}]
\gamma^{\mu}P_{L}\\&&
+[-2m_{t}(C_{21}^{b}+2C_{11}^{b}+C_{0}^{b})]p_{t}^{\mu}P_{L}+[2m_{t}(C_{23}^{b}+2C_{12}^{b})]p_{c}^{\mu}P_{L}\},
\end{eqnarray*}
\begin{eqnarray*}
\Gamma^{\mu}_{tcZ}(\omega^{\pm})&=&\frac{i}{16\pi^{2}}\frac{g}{\cos\theta_{W}}
\frac{g^{2}}{2M_{W_{H}}^{2}}(V_{Hu})_{3i}(V_{Hu})_{i2}{m_{Hi}^{2}}\\&&
\{[(\frac{1}{2}-\frac{2}{3}\sin^{2}\theta_{W})
(B_{0}(-p_{t},m_{Hi},0)
-B_{0}(-p_{c},m_{Hi},0)\\&&+B_{1}(-p_{t},m_{Hi},0))
+(-\frac{1}{2}+\frac{1}{3}\sin^{2}\theta_{W})(2C_{24}^{c}-2p_{t}\cdot
p_{c}
(C_{12}^{c}+C_{23}^{c})\\&&+m_{t}^{2}(C_{21}^{c}+C_{11}^{c}+C_{0}^{c})-m_{Hi}^{2}C_{0}^{c})
+2\cos^{2}\theta_{W}C_{24}^{g}] \gamma^{\mu}P_{L}\\&&
+[(-\frac{1}{2}+\frac{1}{3}\sin^{2}\theta_{W})
(-2m_{t}(C_{21}^{c}+2C_{11}^{c}+C_{0}^{c}))\\&&+\cos^{2}\theta_{W}m_{t}(2C_{21}^{g}+C_{11}^{g})]p_{t}^{\mu}P_{L}
+[2(-\frac{1}{2}+\frac{1}{3}\sin^{2}\theta_{W})
m_{t}(C_{23}^{c}+2C_{12}^{c})\\&&-\cos^{2}\theta_{W}m_{t}(2C_{23}^{g}+C_{11}^{g})]p_{c}^{\mu}P_{L}\},
\end{eqnarray*}
\begin{eqnarray*}
\Gamma^{\mu}_{tcZ}(A_{H})&=&\frac{i}{16\pi^{2}}\frac{g}{\cos\theta_{W}}
(\frac{1}{2}-\frac{2}{3}\sin^{2}\theta_{W})\frac{g^{\prime2}}{50}
(V_{Hu})_{3i}(V_{Hu})_{i2}\\&&
\{[B_{1}(-p_{t},m_{Hi},M_{A_{H}})+2C_{24}^{d} -2p_{t}\cdot
p_{c}(C_{11}^{d}+C_{23}^{d})+m_{t}^{2}(C_{21}^{d}+C_{11}^{d})\\&&-m_{Hi}^{2}C_{0}^{d}]
\gamma^{\mu}P_{L}
+[-2m_{t}(C_{21}^{d}+C_{11}^{d})]p_{t}^{\mu}P_{L}
+[2m_{t}(C_{23}^{d}+C_{11}^{d})]p_{c}^{\mu}P_{L}\},
\end{eqnarray*}
\begin{eqnarray*}
\Gamma^{\mu}_{tcZ}(Z_{H})&=&\frac{i}{16\pi^{2}}\frac{g}{\cos\theta_{W}}
(\frac{1}{2}-\frac{2}{3}\sin^{2}\theta_{W})\frac{g^{2}}{2}
(V_{Hu})_{3i}(V_{Hu})_{i2}\\&&
\{[B_{1}(-p_{t},m_{Hi},M_{Z_{H}})+2C_{24}^{e} -2p_{t}\cdot
p_{c}(C_{11}^{e}+C_{23}^{e})+m_{t}^{2}(C_{21}^{e}+C_{11}^{e})\\&&-m_{Hi}^{2}C_{0}^{e}]
\gamma^{\mu}P_{L}
+[-2m_{t}(C_{21}^{e}+C_{11}^{e})]p_{t}^{\mu}P_{L}
+[2m_{t}(C_{23}^{e}+C_{11}^{e})]p_{c}^{\mu}P_{L}\},
\end{eqnarray*}
\begin{eqnarray*}
\Gamma^{\mu}_{tcZ}(W_{H}^{\pm})&=&\frac{i}{16\pi^{2}}\frac{g}{\cos\theta_{W}}{g^{2}}(V_{Hu})_{3i}(V_{Hu})_{i2}\\&&
\{[(\frac{1}{2}-\frac{2}{3}\sin^{2}\theta_{W})
B_{1}(-p_{t},m_{Hi},M_{W_{H}})\\&&+(-\frac{1}{2}
+\frac{1}{3}\sin^{2}\theta_{W})(2C_{24}^{f}-2p_{t}\cdot
p_{c}(C_{11}^{f}+C_{23}^{f})+m_{t}^{2}(C_{21}^{f}
+C_{11}^{f})-m_{Hi}^{2}C_{0}^{f})\\&&+\frac{1}{6}\cos^{2}\theta_{W}(2B_{0}(p_{c},m_{Hi},M_{W_{H}})
+4C_{24}^{h}-4p_{t}\cdot
p_{c}(C_{11}^{h}+C_{0}^{h})\\&&+m_{t}^{2}(3C_{11}^{h}+C_{0}^{h})
+2M_{W_{H}}^{2}C_{0}^{h})]\gamma^{\mu}P_{L}\\&&+[(-\frac{1}{2}
+\frac{1}{3}\sin^{2}\theta_{W})(-2m_{t}(C_{21}^{f}+C_{11}^{f}))\\&&
+\frac{1}{3}\cos^{2}\theta_{W}m_{t}(2C_{21}^{h}+3C_{11}^{h}+C_{0}^{h})]p_{t}^{\mu}P_{L}\\&&
+[2(-\frac{1}{2}+\frac{1}{3}\sin^{2}\theta_{W})m_{t}(C_{23}^{f}+C_{11}^{f})\\&&
-\frac{1}{3}\cos^{2}\theta_{W}m_{t}(2C_{23}^{h}+3C_{12}^{h}-C_{11}^{h}-C_{0}^{h})]p_{c}^{\mu}P_{L}\},
\end{eqnarray*}
\begin{eqnarray*}
\Gamma^{\mu}_{tcZ}(W_{H}^{\pm}\omega^{\pm})&=&\frac{i}{16\pi^{2}}g\cos\theta_{W}\frac{g^{2}}{2}
(V_{Hu})_{3i}(V_{Hu})_{i2}\\&&\{[m_{Hi}^{2}(C_{0}^{i}-C_{0}^{j})+m_{t}^{2}(C_{11}^{j}+C_{0}^{j})]
\gamma^{\mu}P_{L}+[-2m_{t}C_{12}^{j}]p_{c}^{\mu}P_{L}\}\\
\end{eqnarray*}
Here $i,j$ are the color indexes and $a$ is the index of gluon.
The three-point standard functions $C_0,~C_{ij}$ are defined as
\begin{eqnarray*}
C_{ij}^{a}&=&C_{ij}^{a}(-p_{t},p_{c},m_{Hi},0,m_{Hi}),
\end{eqnarray*}
\begin{eqnarray*}
C_{ij}^{b}&=&C_{ij}^{b}(-p_{t},p_{c},m_{Hi},0,m_{Hi}),
\end{eqnarray*}
\begin{eqnarray*}
C_{ij}^{c}&=&C_{ij}^{c}(-p_{t},p_{c},m_{Hi},0,m_{Hi}),
\end{eqnarray*}
\begin{eqnarray*}
C_{ij}^{d}&=&C_{ij}^{d}(-p_{t},p_{c},m_{Hi},M_{A_{H}},m_{Hi}),
\end{eqnarray*}
\begin{eqnarray*}
C_{ij}^{e}&=&C_{ij}^{e}(-p_{t},p_{c},m_{Hi},M_{Z_{H}},m_{Hi}),
\end{eqnarray*}
\begin{eqnarray*}
C_{ij}^{f}&=&C_{ij}^{f}(-p_{t},p_{c},m_{Hi},M_{W_{H}},m_{Hi}),
\end{eqnarray*}
\begin{eqnarray*}
C_{ij}^{g}&=&C_{ij}^{g}(-p_{t},p_{c},0,m_{Hi},0),
\end{eqnarray*}
\begin{eqnarray*}
C_{ij}^{h}&=&C_{ij}^{h}(-p_{t},p_{c},M_{W_{H}},m_{Hi},M_{W_{H}}),
\end{eqnarray*}
\begin{eqnarray*}
C_{ij}^{i}&=&C_{ij}^{i}(-p_{t},p_{c},M_{W_{H}},m_{Hi},0),
\end{eqnarray*}
\begin{eqnarray*}
C_{ij}^{j}&=&C_{ij}^{j}(-p_{t},p_{c},0,m_{Hi},M_{W_{H}}).
\end{eqnarray*}

\newpage
\begin{figure}[h]
\begin{center}
\epsfig{file=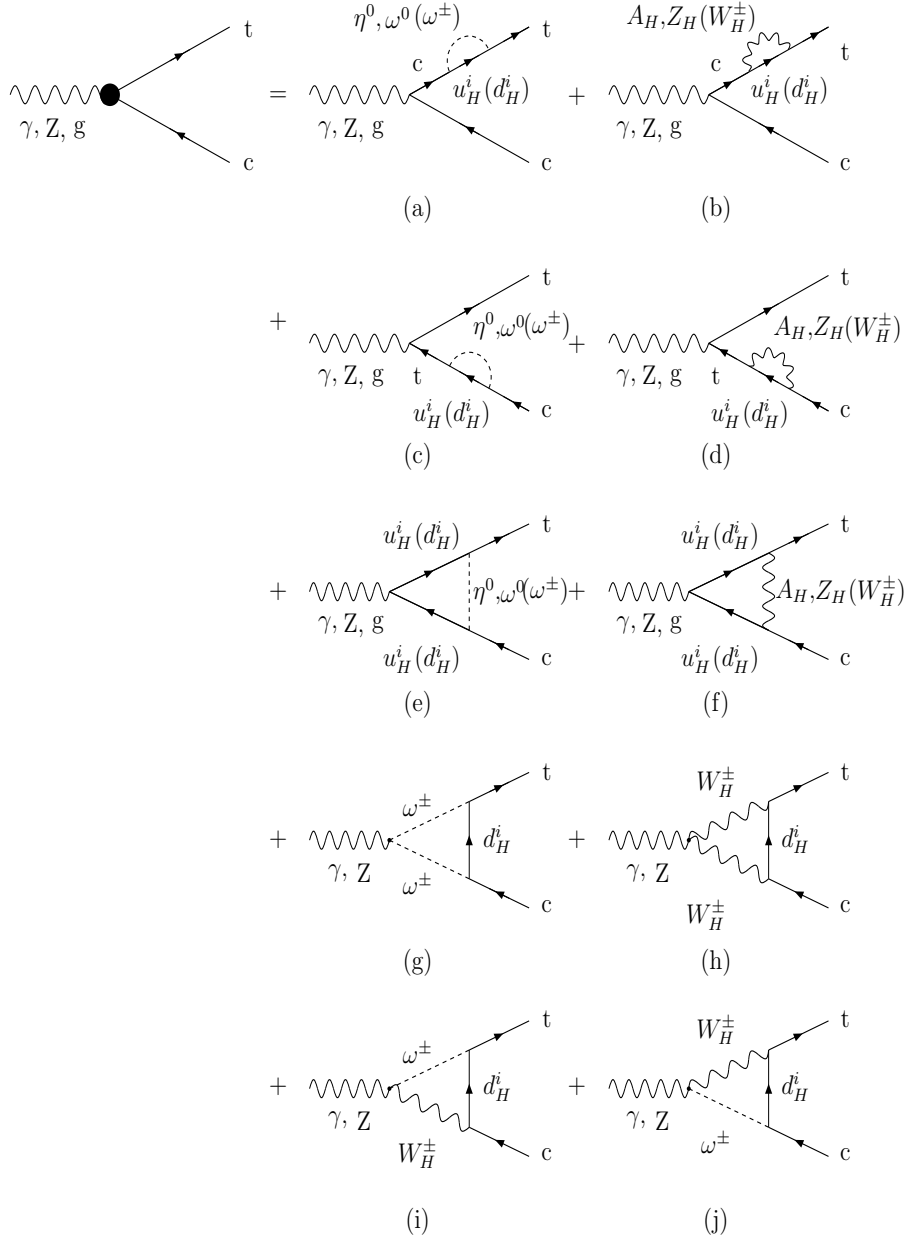,width=450pt,height=720pt} \vspace{-5cm}
\caption{\small One-loop contributions of the LHT model to the
$tcV$ couplings.}
\end{center}
\end{figure}

\newpage
\begin{figure}[h]
\begin{center}
\epsfig{file=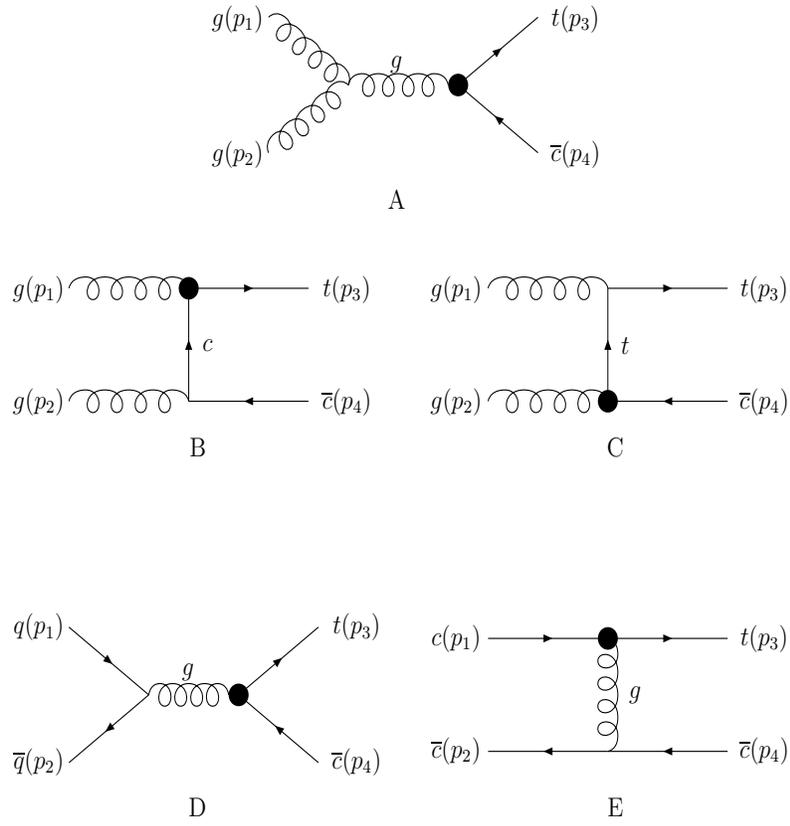,width=450pt,height=720pt} \vspace{-11cm}
\caption{\small The Feynman diagrams of the subprocesses
$gg(q\bar{q})\rightarrow t\bar{c}$ in the LHT model.}
\end{center}
\end{figure}

\newpage

\begin{figure}[h]
\begin{center}
\epsfig{file=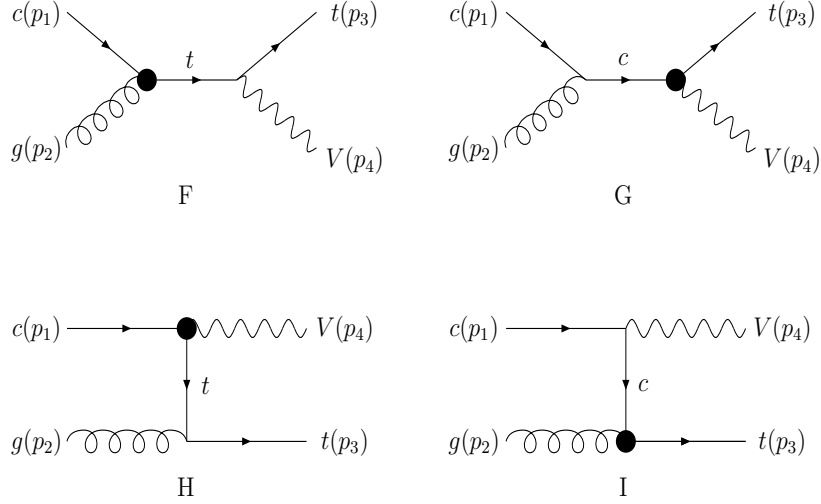,width=450pt,height=720pt} \vspace{-15cm}
\caption{\small The Feynman diagrams of the subprocesses
$cg\rightarrow tV(V=\gamma,Z,g)$ in the LHT model.}
\end{center}
\end{figure}

\newpage

\begin{figure}[h]
\scalebox{0.7}{\epsfig{file=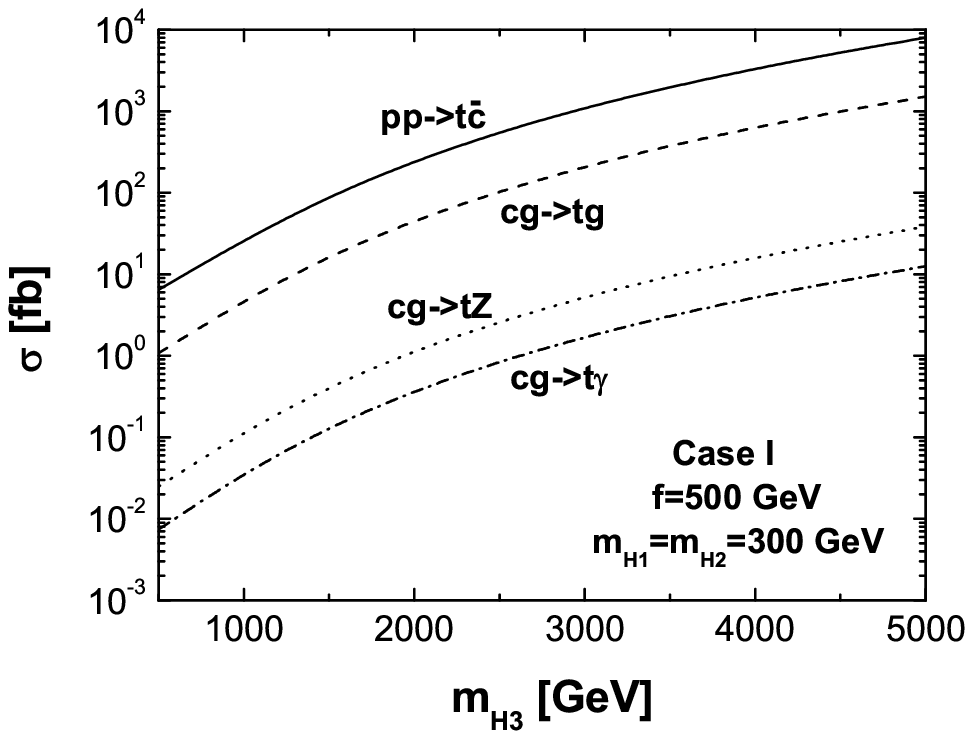}}
\scalebox{0.7}{\epsfig{file=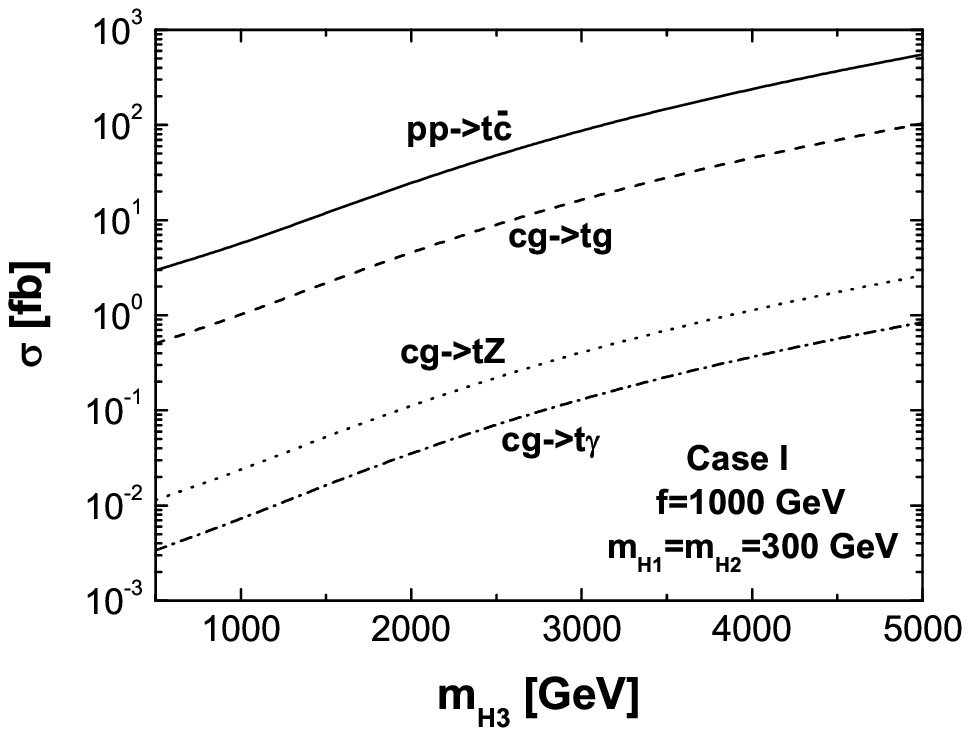}}\\
\caption  {The cross sections of the processes $pp \rightarrow
t\bar{c}$ and $pp \rightarrow tV$ in the LHT model at the LHC for
Case I, as a function of $M_{H_3}$. Here we fix
$m_{H_1}=m_{H_2}=300$ GeV and take f=500 GeV(left figure),f=1000
GeV(right figure), respectively.}
\end{figure}

\newpage

\begin{figure}[h]
\scalebox{0.7}{\epsfig{file=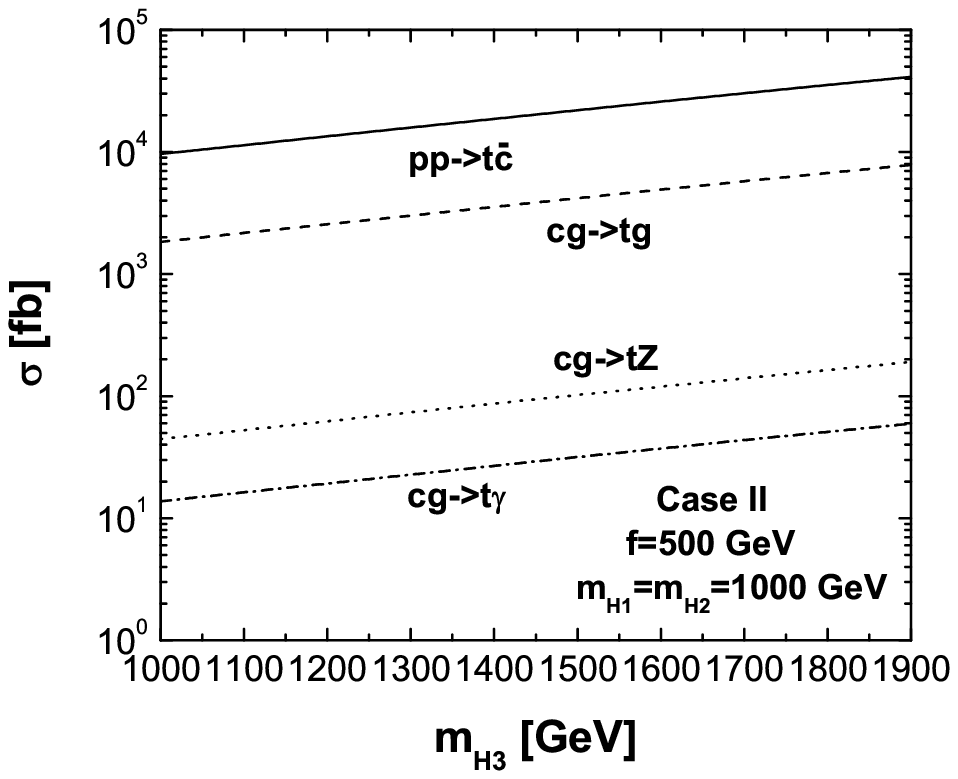}}
\scalebox{0.7}{\epsfig{file=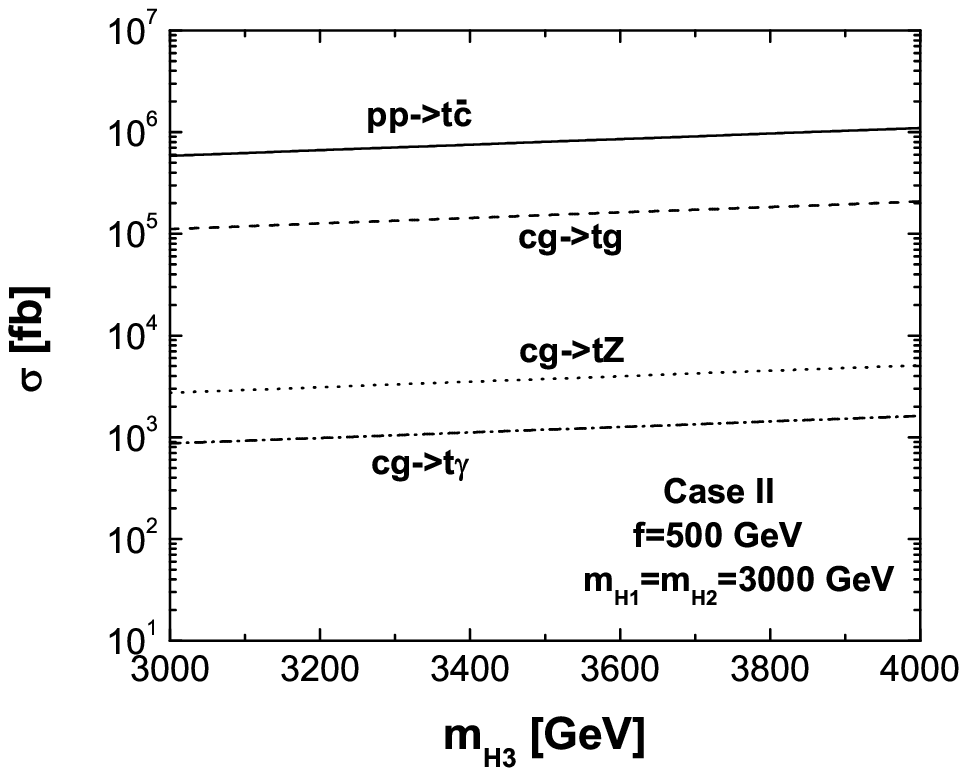}}\\
\caption  {The cross sections of the processes $pp \rightarrow
t\bar{c}$ and $pp \rightarrow tV$ in the LHT model at the LHC for
Case II, as a function of $M_{H_3}$. Here we fix $f=500$ GeV and
take $m_{H_1}=1000$ GeV(left figure), $m_{H_1}=3000$ GeV(right
figure), respectively.}
\end{figure}

\end{document}